\documentclass[%
 reprint,%
 amssymb, amsmath,%
 aip,cha,%
]{revtex4-1}

\usepackage{docs}%
\usepackage{graphicx}
\usepackage{dcolumn}
\usepackage{bm}
\usepackage{units}
\usepackage[colorlinks=true,linkcolor=blue]{hyperref}%
\expandafter\ifx\csname package@font\endcsname\relax\else
 \expandafter\expandafter
 \expandafter\usepackage
 \expandafter\expandafter
 \expandafter{\csname package@font\endcsname}%
\fi
\def\mathclap#1{\text{\hbox to 0pt{\hss$\mathsurround=0pt#1$\hss}}} 
\newcommand{\msub}[1]{_{\text{#1}}}

\DeclareFontFamily{U}{euc}{}
\DeclareFontShape{U}{euc}{m}{n}{<-6>eurm5<6-8>eurm7<8->eurm10}{}%
\DeclareSymbolFont{AMSc}{U}{euc}{m}{n} 
\DeclareMathSymbol{\umu}{\mathord}{AMSc}{"16}

\begin{document}


\title{On the field dependent surface resistance of niobium on copper cavities}

\author{T. Junginger}
 \altaffiliation{Electronic mail: Tobias.Junginger@cern.ch}
 \affiliation{CERN, Geneva, Switzerland}

\date{\today}

\begin{abstract}
The surface resistance $R\msub{S}$ of superconducting cavities prepared by sputter coating a thin niobium film on a copper substrate increases significantly stronger with the applied RF field compared to cavities of bulk material. A possible cause is that due to the thermal boundary resistance between the copper substrate and the niobium film $R\msub{S}$ is enhanced due to global heating of the inner cavity wall. Introducing helium gas in the cavity and measuring its pressure as a function of applied field allowed to conclude that the inner surface of the cavity is heated up by only \unit[60$\pm $60]{mK} when $R\msub{S}$ increases with $E\msub{acc}$ by \unit[100]{n$\Omega$}. This is more than one order of magnitude less than what one would expect from global heating. Additionally the effect of cooldown speed and low temperature baking have been investigated in the framework of these experiments. It is shown that for current state of the art niobium on copper cavities there is only a detrimental effect of low temperature baking. A fast cooldown results in a lowered $R\msub{S}$. 

\end{abstract}

\maketitle
\section{Introduction}
Superconducting cavities prepared by coating a micrometer thin niobium film on a copper substrate are currently used at CERN in the LHC and for the HIE-Isolde project \cite{ISOLDE}. The operation temperature here is \unit[4.5]{K}, where this technology enables a lower surface resistance than cavities prepared from bulk material. Other advantages of Nb/Cu cavities are lower material cost, no need for shielding the earth's magnetic field and enhanced thermal stability avoiding quenches \cite{Calatroni200695}. For quarter-wave resonators cryostat design is facilitated since these copper structures can be cooled by conduction \cite{ISOLDE}. 

Despite these advantages the Nb/Cu technology is currently not considered for accelerators requiring highest accelerating gradient $E\msub{acc}$ like the ILC or lowest surface resistance $R\msub{S}$ at temperatures of \unit[2]{K} and below, especially for high duty cycle and CW applications. The reason is that $R\msub{S}$ increases strongly with $E\msub{acc}$. The origin of this field dependent surface resistance has been subject of many studies in the past \cite{Darriulat,CalatronSRF2001,CalatroniSRF03}, but is still far from being fully understood. No single dominant source can be expected, thus several hypotheses need to be specifically addressed individually to identify their origin and possibly reduce their extent. In this paper the effect of a potential temperature difference between the inner cavity surface and the helium bath, induced by RF heating, is investigated. The influence on the RF performance of the cooldown speed and low temperature baking are addressed. 

An issue in superconducting cavities is
the possibility that the thermal impedance of the
cavity wall leads to an increased temperature of the inner
surface exposed to the RF, resulting in an increased
surface resistance \cite{CalatroniSRF03}. For bulk niobium cavities it has been shown that this effect at least contributes to the field dependent surface resistance \cite{Palmieri2014}. For Nb/Cu cavities the problem is even more complex, since the thermal impedance across the film/substrate interface is unknown and adds to the total thermal conductivity. Here a method previously tried in \cite{CalatroniSRF03} to directly measure the inner surface temperature of a superconducting cavity is studied in detail. 

Thermal cycling has been studied in detail by several laboratories on bulk niobium \cite{Vogt_PRSTAB13,RomanenkoJAP14} and also on nitrogen doped niobium cavities \cite{RomanenkoJAP14}. It is however still unclear, what the ideal cooldown procedure is and if it depends on the material treatment of the cavity. Another open question is how much magnetic flux created by thermal currents influences $R\msub{S}$, which could be the reason for differences observed between vertical and horizontal test results \cite{Vogt_PRSTAB13}. Studies on Nb/Cu cavities are not only useful to push their performance, but also to obtain further information about the mechanism of thermal cycling in general, since Nb/Cu cavities are a lot less sensitive to ambient magnetic fields. 
At CERN such studies have been carried out using the Quadrupole Resonator \cite{Mahner:611593,Junginger:Revscientinstr12} on a Nb/Cu sample prepared by electron cyclotron resonance (ECR) \cite{Aull_TSRF14} and HIE Isolde Nb/Cu quarter wave resonators \cite{Zhang_TSRF14}.
%
In this paper the first study on elliptical Nb/Cu cavities is presented. Results will be compared with the references given above and an advice for a cooldown procedure of cryomodules equipped with Nb/Cu cavities will be given.

  %
%
%

Low temperature in-situ baking is used for bulk niobium cavities to shift the so called Q-drop (exponential increase of the surface resistance at high field) to larger values enabling economical high gradient operation \cite{Ciovati_JAP04}. In the medium field region above \unit[10]{MV/m} Nb/Cu cavities also exhibit an exponential increase of $R\msub{S}$ with $E\msub{acc}$. One might therefore argue that the origin for this is the same as for the Q-drop of bulk niobium cavities. To test this hypothesis the cavity used in this study was tested under same conditions before and after a $150\,^{\circ}{\rm C}$ bake-out. 
\section{Experimental setup and procedure}
\begin{figure}[tbp]
   \centering
	 \includegraphics[width=\columnwidth]{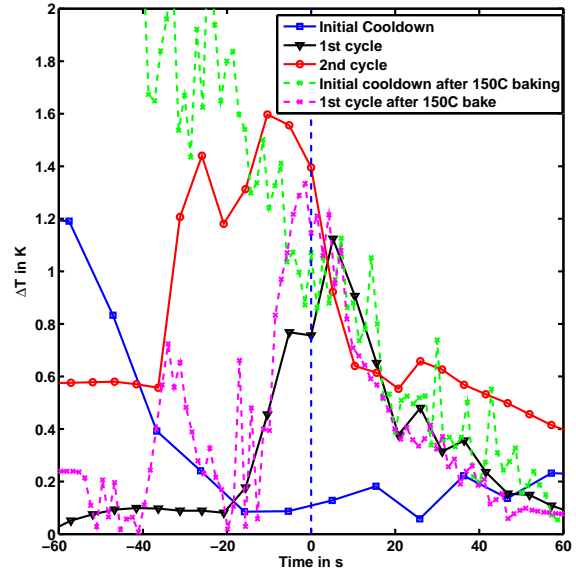}
   \caption{Temperature gradient along the cell from one minute before until one minute after the phase transition for five cooldowns.}
   \label{figure:cooldown}
\end{figure}

A micrometer thin niobium film has been deposited by High-power Impulse Magnetron Sputtering (HIPIMS) technology on the inner surface of an elliptical single-cell TESLA type\cite{PhysRevSTAB.3.092001} \unit[1.3]{Ghz} copper cavity produced by spinning and electopolishing. The important cavity parameters for the studies presented here are found in Tab. \ref{tab:TESLA}. They relate the peak magnetic and electric fields to the accelerating gradient $E\msub{acc}$ and the unloaded quality factor $Q\msub{0}$ to the surface resistance $R\msub{S}$. In the following always $E\msub{acc}$ and $R\msub{S}$ will be used. 
\begin{table}[b]
   \centering
  \caption{Parameters of TESLA type cavites\cite{PhysRevSTAB.3.092001}}
   \begin{tabular}{|l|c|}
			\hline
      Geometry Factor G &  270 $\Omega$ \\      
				\hline
      $E\msub{peak}/E\msub{acc}$ & 2.0 \\
			\hline
			$B\msub{peak}/E\msub{acc}$ & \unit[4.26]{mT/(MV/m)}\\
			\hline
			\end{tabular}
  \label{tab:TESLA}
\end{table}
The HIPIMS approach is under investigation at CERN since 2010 \cite{TerenzianiSRF13}. So far the results show systematically an increase of $R\msub{S}$ with $E\msub{acc}$ similar to what has been observed in the past for cavities coated by dc magnetron sputtering (dcms) \cite{Darriulat}. For the vertical cold tests the cavity has been equipped with an input coupler with a $Q\msub{ext}=2\times 10^9$ and a pick-up antenna with $Q\msub{pick-up}=3\times 10^{11}$. The external coupling $Q\msub{ext}$ was chosen to obtain critical coupling at \unit[1.8]{K} and $E\msub{acc} \approx$\unit[15]{MV/m}. RF measurements were performed using the phase locked loop technique \cite{Padamsee:1116813}. Temperature sensors were mounted above and below the cavity cell to measure the temperature and its gradient during the phase transition from the normal to the superconducting state. The vertical bath cryostat has been equipped with a mu-metal shield to reduce the ambient magnetic field to about \unit[3]{$\mu$T}. Helium was transferred using standard dewars pressurized with helium gas. For the initial cooldown the gas overpressure was set to about \unit[200]{mbar}. During this first helium transfer the dewar was emptied when the cavity temperature was about \unit[50]{K}. This lead to a thermalization of the cavity. The temperature gradient at the phase transition was the lowest of all cooldowns performed, see Fig. \ref{figure:cooldown}.

In total the cavity was cooled down five times. Each cycle started with a measurement of $R\msub{S}$ as a function of $E\msub{acc}$ at \unit[4.2]{K}. Then the cavity was further cooled down to \unit[1.8]{K} by pumping the helium bath to \unit[16.5]{mbar}. This process takes several hours, which is slow enough to take several measurements of $R\msub{S}$ at $E\msub{acc}$=\unit[1]{MV/m}. After the initial cooldown the cavity was warmed up twice above $T\msub{c}$ and cooled down again. During each warmup the quality factor was measured with a network analyzer to ensure that the whole cavity was in the normal conducting state before the helium transfer was restarted. The maximum temperature the cavity reached in these cycles never exceeded \unit[20]{K}. For the first thermal cycle the overpressure was set relatively high to about \unit[250]{mbar} on the helium supply dewar, while for the second cycle the overpressure was set to only \unit[50]{mbar} to reduce the cooldown speed. Figure \ref{figure:cooldown} displays the temperature difference between the upper and the lower sensor from one minute before until one minute after the phase transition for the three cycles. The cooldown speed was about \unit[70]{mK/s} for the initial cooldown and the second cycle and \unit[230]{mK/s} for the first cycle.
%
%

\begin{figure}[tbp]
   \centering
	 \includegraphics[width=\columnwidth]{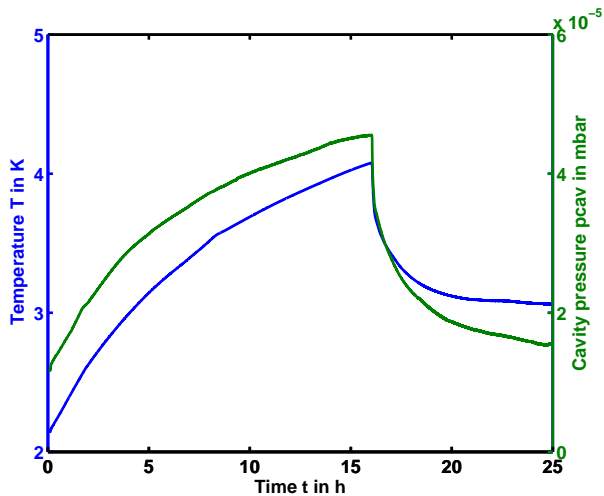}
   \caption{Temperature and pressure in the cavity during warmup and following cooldown measured as a function of time.}
   \label{figure:p_T_vs_t}
\end{figure}

\begin{figure}[tbp]
   \centering
	\includegraphics[width=\columnwidth]{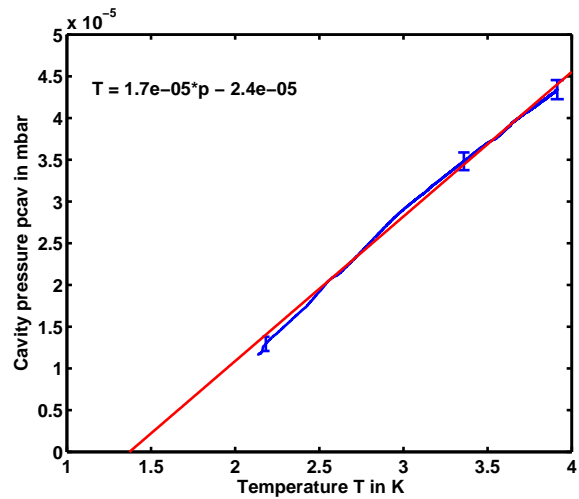}
   \caption{Cavity pressure as a function of temperature measured during warm-up. The blue line represents measurements obtained every \unit[10]{s}, while the red line shows a linear fit to the data.}
   \label{figure:pvsT}
\end{figure}

After the initial cooldown several additional scans of $R\msub{S}$ vs $E\msub{acc}$ were performed just below and above the lambda point of helium at \unit[2.17]{K}. During the warmup to room temperature the frequency shift was measured as a function of temperature to derive the penetration depth. For this the helium was almost completely evaporated so that the whole cavity was only cooled by helium gas. A few liters left on the bottom allowed to obtain a constant temperature along the cavity by eva\-po\-ra\-ting some of the liquid. For a precise measurement pressure control is extremely crucial. A PID-controlled butterfly valve was used to pump the bath pressure to \unit[80]{mbar}. A stability better than \unit[0.02]{mbar} was achieved during the whole measurement. After complete warmup the cavity was in situ baked at $150\,^{\circ}{\rm C}$ for 50 h. In the following RF tests only one thermal cycle with a fast cooldown has been carried out. The speed was \unit[110]{mK/s} for the initial cooldown and \unit[330]{mK/s} for the thermal cycle, after which an additional measurement was performed. Helium gas was introduced in the cavity. At \unit[1.8]{K} its pressure was $p\msub{RT}$=\unit[5$\times 10^{-5}$]{mbar} measured at the top flange connection of the cryostat, which corresponds, according to the thermal transpiration law 
\begin{equation}
p\msub{cav}=p\msub{RT}\sqrt{\frac{T\msub{cav}}{T\msub{RT}}}
\end{equation}
to a cavity pressure $p\msub{cav}$=\unit[4$\times 10^{-6}$]{mbar}, where $T\msub{cav}$ and $T\msub{RT}$ are the cavity and the room temperature, respectively. The surface resistance was once again measured as a function of $E\msub{acc}$, now with the helium gas inside the cavity. To test whether the temperature of the cavity's inner surface increases with $E\msub{acc}$ the helium pressure was recorded. In order to calibrate the pressure to temperature values during the following warm-up both quantities were recorded. A clear correlation was found, see Fig. \ref{figure:p_T_vs_t}. To verify that the pressure increase was not caused by a leak introducing just more helium gas with time the helium bath was pumped down again. Once more a clear correlation between $p$ and $T$ was found, see Fig. \ref{figure:p_T_vs_t}.
For the cooldown lower pressures were measured for the same temperatures, which can be correlated to the lower helium level in the cryostat. The gas in the now warmer region of the pumping line is expanding more and contributes to the total pressure. For the calibration of $p$ with respect to $T$ only the values obtained during warm-up were used. These have been recorded at almost the same liquid helium level as the $R\msub{S}$ and $p$ vs. $E\msub{ac}$ measurement directly before. The data and a linear fit are displayed in Fig. \ref{figure:pvsT}. The physical mechanism, which is responsible for the pressure increase as a function of temperature involves temperature dependent adsorption on the cavity wall and expansion of the non-adsorbed gas with temperature. The pressure of $p\msub{cav}$=\unit[4$\times 10^{-6}$]{mbar} was chosen to be high enough that the surface was saturated with helium and a temperature increase results mainly in an increase of the pressure of the non-absorbed gas. Based on data from \cite{Benvenuti84} it can be estimated that at the pressure used only a fraction of about \unit[10]{\%} can be absorbed by the cold cavity surface. An even higher pressure has been tried but resulted in strong discharging at a few MV/m making the measurement impossible.
Saturation of the cold surface is important when the pressure increase of the global warming of the cavity due to warming of the helium bath for calibration is compared with the pressure increase due to heating from RF. For example if one would heat up by RF only the cavity cell releasing gas from its surface it could be adsorbed by the non-saturated cut-off tubes remaining at bath temperature. In this case a warming of the cavity cell could mainly result in an increase of the adsorbed gas on the cut-off tube surfaces. The linear dependence measured between $p$ and $T$ (Fig. \ref{figure:pvsT}) suggests that the ideal gas equation is indeed applicable and the surfaces are well saturated.   
%
%

\section{Results}
\subsection{Low field surface resistance} 
%
%
 
\begin{figure}[tbp]
   \centering
	\includegraphics[width=\columnwidth]{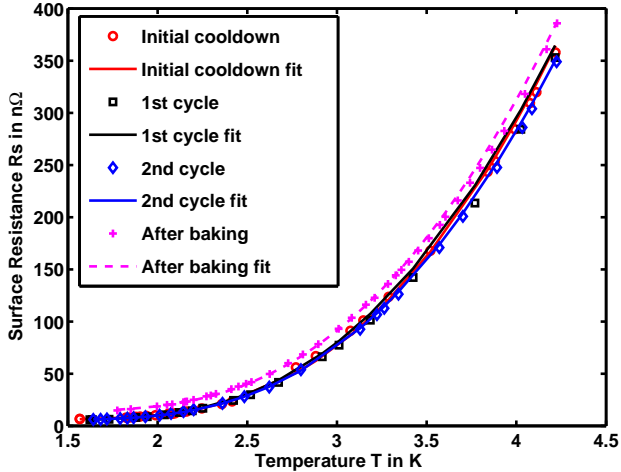}
   \caption{Surface resistance $R\msub{S}$ of a \unit[1.3]{GHz} Nb/Cu cavity as a function of temperature measured after initial cooldown and two following thermal cycles above $T\msub{c}$.}
   \label{figure:RvsT_cycling}
\end{figure}

\begin{figure}[tbp]
   \centering
	\includegraphics[width=\columnwidth]{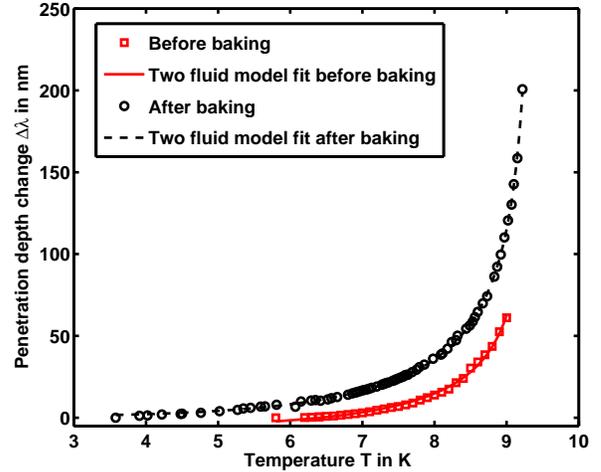}
   \caption{Penetration depth change as a function of temperature measured before and after baking.}
   \label{figure:PenDepth}
\end{figure}

%
For each cooldown cycle the surface resistance was measured as a function of temperature at $E\msub{acc}$=\unit[1]{MV/m}, see Fig. \ref{figure:RvsT_cycling}. The data has been fitted to BCS theory using the WinSuperFit software based on calculations from Halbritter \cite{Ciovati_Win_super_Fit_old}. As input parameters the critical temperature $T\msub{c}$=\unit[9.41]{K}, the London penetration depth for infinite mean free path $\lambda\msub{L}$=\unit[32]{nm} and the BCS coherence length $\xi_0$=\unit[39]{nm} were used. The former quantity was obtained from frequency shift measurements (see below), while the latter two were taken from the literature \cite{MattisBardeenTheory}. Three parameters were varied in a least squares fit, see Tab. \ref{tab:Fit_Parameters}. The smallest residual resistance $R\msub{res}$ was obtained for the fastest cooldown. The effect is however relatively small. 
The low temperature bake-out yielded an increase of $R\msub{res}$ by several n$\Omega$. The mean free path could not be accurately determined from $R\msub{S}$ vs. $T$, because it was not known whether the mean free path is larger or shorter than the coherence length. For thin film cavities this is not obvious, especially not after low temperature baking. From a measurement of the frequency shift, see Fig \ref{figure:PenDepth} however one can determine $l$ independently even though it has to be noted that here the information depth is different (several \unit[100]{nm}) than for the $R\msub{S}$ vs. $T$ measurement  (about \unit[50]{nm}). From a least squares fit to the Gorter-Casimir expression \cite{Gorter1934306}
\begin{equation}
\lambda(T)=\frac{\lambda_0}{\sqrt{1-(\frac{T}{T\msub{c}})^4}}
\end{equation} 
$\lambda_0$ and $T\msub{c}$ are derived. From $\lambda_0$ the electron mean free path $l$ can be obtained from an expression found by Pippard \cite{1953}:
\begin{equation}
\lambda(l)=\lambda(l\rightarrow \infty)\sqrt{1+\frac{\pi\xi_0}{2l}}.
\end{equation}
The London penetration depth $\lambda_L$=\unit[32]{nm} and the BCS coherence length $\xi_0$=\unit[39]{nm} were taken once more from literature \cite{MattisBardeenTheory}. This measurement shows that the low temperature baking significantly decreases $l$, see Tab. \ref{tab:Fit_Parameters}. The BCS surface resistance is only slightly altered because it has its minimum at $l$=$\xi_0$ \cite{MattisBardeenTheory}.
%
%
%
%
%
%
\begin{table*}[t]
   \centering
  \caption{Fit parameters obtained from low field surface resistance and penetration depth change measurement}
   \begin{tabular}{l|cccc}
        &initial cooldown & 1st cycle & 2nd cycle & after baking  \\        
       \hline
       $\Delta/(k\msub{b}T\msub{c})$ &1.96$\pm$0.03 &1.90$\pm$0.05 & 1.93$\pm$0.01& 1.89$\pm$0.01\\
			$R\msub{Res}$ in n${\Omega}$ &6.1$\pm$0.2 & 5.2$\pm$0.2& 5.5$\pm$0.2& 13.7$\pm$0.4\\
			$l$ in nm from R vs T &85$\pm$40 &25$\pm$120&25$\pm$120 & 27$\pm$130 \\
			\hline
			$\lambda\msub{0}$ in nm & & & 46$\pm$4& 78$\pm$1 \\
			$l$ in nm from $\Delta\lambda$& & &58$\pm$4 & 12$\pm$1\\
			$T\msub{c}$ & -& - & 9.41$\pm$0.05 & 9.41$\pm$0.02 \\
   \end{tabular}
   \label{tab:Fit_Parameters}
\end{table*}

\subsection{Field dependent surface resistance} 

\begin{figure}[tbp]
   \centering
	\includegraphics[width=\columnwidth]{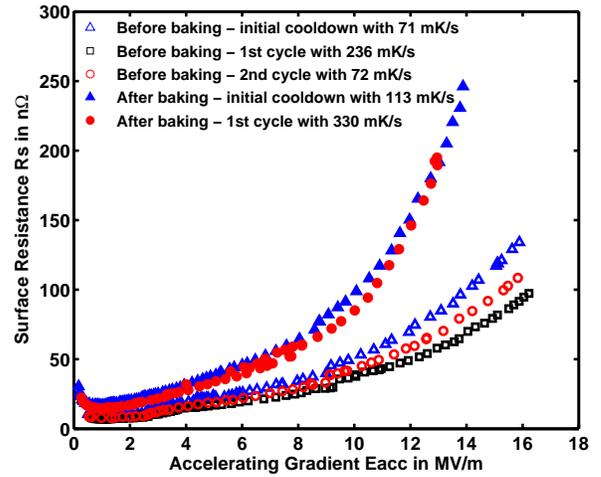}
   \caption{Surface resistance $R\msub{S}$ of a \unit[1.3]{GHz} Nb/Cu cavity measured at \unit[1.8]{K} before (open symbols) and after baking (closed symbols) for initial cooldown and several thermal cycles above $T\msub{c}$.}
   \label{figure:1_8_cycles}
\end{figure}

Figure \ref{figure:1_8_cycles} shows $R\msub{S}$ as a function of $E\msub{acc}$ at \unit[1.8]{K} for five cooldown cycles, including two after baking. At low field the surface resistance differs only by a few n$\Omega$ as one can also see from Fig. \ref{figure:RvsT_cycling}. At larger values of $E\msub{acc}$ the differences in $R\msub{S}$ become more pronounced. The lowest value is obtained for the fastest cooldown (1st cycle) before baking. Baking enhances $R\msub{S}$ significantly, especially at higher field. After baking $R\msub{S}$ is again reduced by fast cooldown, even though the effect is less pronounced as before. 
   
\begin{figure}[tbp]
   \centering
	\includegraphics[width=\columnwidth]{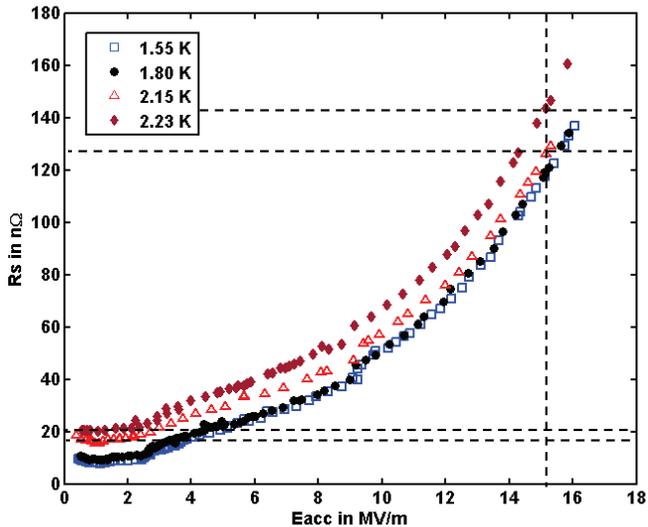}
   \caption{Surface resistance $R\msub{S}$ of a \unit[1.3]{GHz} Nb/Cu cavity measured at four different temperatures. At low field $R\msub{S}$ is similar for the four temperatures. At higher field $R\msub{S}$ increases more strongly if data is taken at temperatures above the lambda point. Compare the horizontal dashed lines.}
   \label{figure:RvsE}
\end{figure}

After the initial cooldown $R\msub{S}$ was additionally measured as a function of $E\msub{acc}$ for several temperatures just above and below the lambda point of helium at \unit[2.17]{K}, see Fig. \ref{figure:RvsE}. One can see that at low field $R\msub{S}$ varies only slightly for the different temperatures. However for higher accelerating gradients $R\msub{S}$ increases stronger with $E\msub{acc}$ if the helium bath is not superfluid. Between 1.55 and \unit[2.13]{K} the temperature dependent BCS surface resistance changes by a factor of about 15 from about 1 to approx \unit[15]{n$\Omega$}, see Fig. \ref{figure:RvsT_cycling}. The increase of $R\msub{S}$ with $E\msub{acc}$ on the other hand is very similar for the curves of 1.55, 1.8 and \unit[2.13]{K}, see Fig. \ref{figure:RvsE}. This shows that the field dependent surface resistance of Nb/Cu cavities is at least dominated by the residual resistance.    

\begin{figure}[tbp]
   \centering
	\includegraphics[width=\columnwidth]{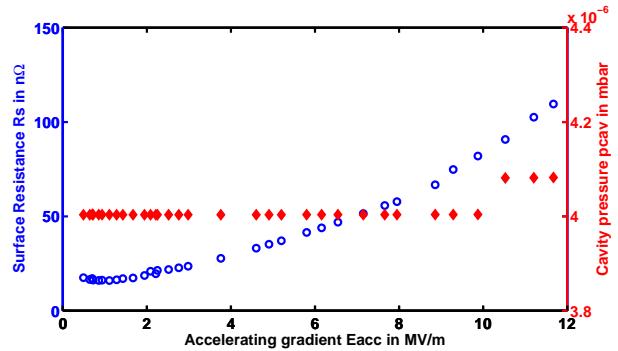}
   \caption{Primary y-axis: Surface resistance $R\msub{S}$ of a \unit[1.3]{GHz} Nb/Cu cavity measured at \unit[1.8]{K}. Secondary y-axis: Helium pressure $p$ inside the cavity as a function of $E\msub{acc}$.}
   \label{figure:p_and_R_vs_E}
\end{figure}

After baking and the following thermal cycle the cavity was filled with helium gas. Then $R\msub{S}$ vs. $E\msub{acc}$ was measured again and also the helium pressure was recorded, see Fig. \ref{figure:p_and_R_vs_E}. The resolution of the pressure sensor was only 2 digits which resulted with the calibration obtained during the following warm-up (Fig. \ref{figure:pvsT}) in a resolution of \unit[60]{mK}. Above \unit[10]{MV/m} a pressure increase by \unit[0.1$\times 10^{-6}$]{mbar} was detected, corresponding to a temperature increase of \unit[60$\pm$60]{mK}.

%
%


\section{Discussion}
\subsection{Thermal Cycling}
It was shown that fast cooling can reduce the surface resistance of elliptical Nb/Cu cavities by about 20 \% at $E\msub{acc}$=\unit[15]{MV/m} in a vertical test cryostat. Thermal cycling changes the low field residual resistance only slightly by fractions of a n$\Omega$. The benefit is only at higher accelerating gradients, similar to what has been observed in \cite{RomanenkoJAP14} for low temperature baked 9-cell bulk niobium cavities. However for the Nb/Cu cavity investigated here this effect was reduced by baking. Furthermore it has to be noted that the cooldown speed was at maximum \unit[30]{mK/s} in \cite{RomanenkoJAP14} which is not even half as fast as the lowest cooldown speed of \unit[70]{mK/s} obtained in the experiments here.  In \cite{Vogt_PRSTAB13} it took at least one minute between the transition from the coldest to the warmest part of the cavity, here this was at maximum a few seconds. In \cite{Aull_TSRF14} it was found by measurements using the CERN Quadrupole Resonator on a Nb/Cu sample prepared by electron cyclotron resonance (ECR) that a fast cooldown decreases $R\msub{S}$. The cooldown speed obtained in this experiment was comparable to what has been reported here. However it was found in contradiction to the results presented here that mainly the low field $R\msub{S}$ is affected, while the influence on the field dependence is rather small. The results presented here suggest that the thermal gradient does not affect $R\msub{S}$ but the cooldown speed is of importance, which is in contradiction to what has been measured on Nb/Cu quarter wave resonators in vertical tests\cite{Zhang_TSRF14}. To conclude it can be said that the cooldown speed and the thermal gradient play an important role for minimizing losses in superconducting cavities. Comparing results from different publications is difficult because it is still unclear whether flux trapping from ambient fields or thermal currents are more relevant. Another open question is whether cavities of different geometry and/or different materials/treatments need specific cooldown procedures. Based on the results presented here and taking into account that thermal currents might be more relevant in horizontal cryomodules than in vertical test cryostats the following procedure for horizontal SRF cryomodules equipped with Nb/Cu cavities is advised:
\begin{itemize}
	\item{After the initial cooldown a warmup just above $T\msub{c}$,}
	\item{Give the cavity time to thermalize to minimize thermal currents,}
	\item{Cooldown the cavity as fast as possible through $T\msub{c}$.}
\end{itemize}
This procedure is consistent with all experimental results on Nb/Cu cavities. In \cite{Vogt_PRSTAB13} it was suggested to either thermalize the cavity above $T\msub{c}$ during the first cooldown or perform a thermal cycle. The former option is not advised for Nb/Cu cavities. The thermalized cavity in the first initial cooldown gave an even worse result than for the slowest cycle afterwards.

\subsection{Low temperature baking}
In situ baking of the cavity yielded a higher low field residual resistance and an enhanced field dependent surface resistance, while the temperature dependent BCS surface resistance $R\msub{BCS}$ was almost unaffected. The former two effects have also been observed for bulk niobium cavities \cite{Ciovati_JAP04}. For bulk niobium cavities $R\msub{BCS}$ is lowered by baking, which can be correlated to a decreased mean free path $l$. An unbaked bulk niobium cavity has $l>>\xi_0$. Baking pushes $l$ closer to $\xi_0$. For Nb/Cu cavities baking also reduced $l$, but since these cavities already have $l \approx \xi_0$ before baking there is no reduction of $R\msub{BCS}$. For bulk niobium the low temperature bake is beneficial because it shifts the exponential increase of the surface resistance referred to as Q-Drop to higher fields. Since the baking did not have any similar effect for the Nb/Cu cavity it can be concluded that the origin of the field dependent surface resistance here is a different one than for the Q-drop of bulk niobium cavities. Currently there is no benefit from in situ baking of Nb/Cu cavities. It might however become useful if the medium field Q-slope could be understood and cured and the Q-drop as observed for bulk niobium cavities would become the limitation of this technology. 

\subsection{Thermal resistance} 
Measurements of $R\msub{S}$ as a function of $E\msub{acc}$ just below and above \unit[2.17]{K} the lambda point of helium have shown that the thermal resistance plays a role for the losses in Nb/Cu cavities cooled by helium-I. At \unit[15]{MV/m} the surface resistance is about \unit[20]{n$\Omega$} higher if the cavity is only cooled by helium-I. In helium-II however the contribution to the overall field dependent surface resistance seems to be rather low. An increase of the inner surface temperature of only \unit[60$\pm$60]{mK} was measured. It has to be noted that for the calibration of the gas pressure with respect to the temperature the whole cavity including a part of the pumping line was warmed up and for the RF measurement only the part of the cell with high surface magnetic field. The ratio of these surface areas is roughly two. Taking this into account the heating of the inner surface can at maximum account for an additional $R\msub{S}$ of \unit[6]{n$\Omega$} assuming low field BCS losses. On the other hand an increase of \unit[100]{n$\Omega$} was measured between low field and \unit[15]{MV/m}. It can be concluded that the thermal resistance can play an important role in minimizing the losses of superconducting cavities at high field. The thermal boundary resistance between the Nb film and the copper substrate is however not the main contribution for the field dependent losses of Nb/Cu cavities. In fact the thermal resistance should play a more important role for bulk niobium cavities. In \cite{CalatroniSRF03} it has been reported that the overall thermal conductance of a niobium film on a copper substrate is still four times higher than of electropolished niobium. Even for bulk niobium cavities operated in helium-I it had been concluded by thermal modeling that heating of the inner surface only plays an insignificant role \cite{Geng:587951}. 

The method to measure the gas pressure as a function of the accelerating gradient allowed to derive the inner surface temperature and to calculate an upper estimate how much the heating of the inner surface accounts for the field dependent surface resistance. The experiments presented here were limited by the two digit resolution of the pressure sensor and therefore only allowed to conclude that global heating is not the main contributor to the field dependent surface resistance of Nb/Cu cavities. Further experiments with enhanced resolution and also on bulk niobium cavities are planned to study this effect in more detail. The method has already been tested on an LHC crab cavity and a significant pressure increase was detected. These results will be presented elsewhere. The method might also be used to obtain the quality factor $Q\msub{0}$ in horizontal cryomodule tests equipped with strongly coupled input antennas. In this case one could first measure $Q\msub{0}$ vs $E\msub{acc}$ and $p$ in a vertical test cryostat. After installation in the cryomodule one could insert gas in the cavity again and measure $E\msub{acc}$ and $p$ again and then obtain $Q\msub{0}$ by comparison with the results from the vertical test. This could serve as an independent method to the usual approach of obtaining $Q\msub{0}$ from the liquid helium consumption.   

\section{Conclusion}
It has been shown that a fast cooldown can significantly reduce the surface resistance of Nb/Cu cavities. There is currently no benefit from low temperature baking for this technology. The thermal resistance plays only a minor role for the field dependent losses if the cavity is cooled by helium-II. It can be concluded that the field dependent surface resistance of Nb/Cu is mainly limited by effects intrinsic in the film, which need to be investigated in detail if the current limitations shall be overcome.

\section{Acknowledgment}
I would like to thank Giovanni Terenziani, Serge Forel and Damiano Sonato for preparing the cavity used in this study. Thanks for the technical support from CERN's cryolab staff for the cavity test. The fruitful discussion with my colleagues at CERN, especially Torsten Koettig, Sergio Calatroni, Giovanna Vandoni and Sarah Aull are highly appreciated.

%

\end{document}